\documentclass[journal,twocolumn]{IEEEtran}
\usepackage[utf8]{inputenc}
\usepackage{tabularx}
\usepackage{algorithm}
\usepackage{algpseudocode}
\usepackage{array}
\usepackage{amsfonts}
\usepackage{multirow}
\usepackage{amssymb}
\usepackage{amsmath, amsthm}
\usepackage{mathtools}
\usepackage{bbm}
\usepackage{graphicx}
\usepackage{epstopdf}
\newcommand{\argmin}{\mathop{\mathrm{argmin}}}
\usepackage{epsf}
\usepackage{amsmath, amsfonts}
\usepackage{latexsym}
\usepackage{float}
\usepackage{subcaption}
\usepackage{caption}
\usepackage{multirow}
\usepackage{amssymb}
\usepackage{float}
\usepackage{enumerate}
\usepackage{cite}
\usepackage{color}
\usepackage{scrextend}
\usepackage{bbm}
\usepackage{lipsum}
\usepackage{mathtools}
\usepackage{cuted}
\usepackage{scalerel}
\usepackage{enumitem}
\newtheorem{prop}{Proposition}
\newtheorem{lemma}{Lemma}

\newtheorem{assumption}{Assumption}
\allowdisplaybreaks

\begin{document}
\bstctlcite{IEEEexample:BSTcontrol}

\title{ Enabling Multicast Transmission for Spatio-Temporally Asynchronous User Requests in Wireless Environments}

\author{
Hojung~Lee,~\IEEEmembership{Student Member,~IEEE},
Jun-Pyo~Hong,~\IEEEmembership{Member,~IEEE,},
and~Wan~Choi,~\IEEEmembership{Fellow,~IEEE}
\thanks{H. Lee is with School of Electrical Engineering, Korea Advanced Institute of Science and Technology (KAIST), Daejeon 34141, Korea (e-mail: hojung$\_$lee@kaist.ac.kr).}
\thanks{Jun-Pyo Hong is with the Department of Information and Communications Engineering, Pukyong National University, Busan 48513, South Korea (e-mail: jp\_hong@pknu.ac.kr).}
\thanks{W. Choi is with the Department of Electrical and Computer Engineering and the Institute of New Media and Communications, Seoul National University (SNU), Seoul 08826, Korea (e-mail:wanchoi@snu.ac.kr).}
\thanks{\emph{(Corresponding authors: Jun-Pyo Hong; Wan Choi.)}}
}

\maketitle

\begin{abstract}
The surge in wireless devices and data traffic volume necessitates more efficient transmission methods. Multicasting has garnered consistent attention as a means to fulfill the increasing demand for more efficient data transmission methods. Nevertheless, leveraging multicast wireless networks for spatio-temporally asynchronous data requests poses challenges. In this context, this paper introduces a new multicast mechanism called \emph{set-up based merged multicast (SMMC)} to minimize the delivery time of the requested file in wireless networks by considering the uncertainties inherent in wireless channels. The proposed mechanism comprises two phases. The first phase involves gathering asynchronous requests for a file from users experiencing diverse channel conditions. During this phase, packets of the requested file are transmitted individually in unicast mode within a specified set-up time. Following this, the second phase initiates multicast transmission, which sequentially handles the remaining packets of the file in multicast mode.
In the proposed mechanism, we optimize the set-up time and transmission rates of both unicast and multicast modes to minimize the expected file delivery time by jointly taking into account the statistical characteristics of wireless channels, users' locations, and file popularity. Additionally, we also delve into a \emph{fine-tuned SMMC} by utilizing posterior information on the multicast group size and further improve the performance. Our performance evaluations reveal that the proposed SMMC outperforms conventional unicast methods, especially with high-demand data.
 \end{abstract}


\section{Introduction}
The emergence of the 5G era has ushered in substantial advancements in wireless communication technology, opening the doors to new applications and services spanning across a multitude of industries\cite{Andrews2014}. The evolution of these applications and services has empowered the utilization of numerous wireless communication devices.  Notably, the global number of mobile subscribers is projected to rise from 8.46 billion in 2023 to 9.21 billion by 2029. Concurrently, data traffic per device is expected to surge by over 50 percent \cite{Ericsson2024}.  These trends underscore the growing demand for data communication services and connectivity, highlighting the imperative for faster and more frequent data transmission.

Meanwhile, multicasting plays a crucial role in addressing mobile traffic challenges \cite{Araniti2017}. It enables the simultaneous transmission of data to multiple recipients, increasing bandwidth utilization and conserving resources.\cite{Sun2020}. By eliminating the necessity for separate unicast transmissions, multicasting amplifies transmission efficiency, thereby enhancing the user experience for real-time applications \cite{Seo2021,Seo2022}. This cost-effective solution alleviates the burden on network infrastructure and reduces operational expenses, making multicasting a highly effective approach for managing mobile traffic.
Nevertheless, multicasting requires synchronization among recipients to ensure concurrent reception. This fundamental structure hinders to implement multicasting within asynchronous scenarios where data requests are not aligned in time.

Several studies were conducted to address the asynchronous data requests in multicast transmission. The authors in \cite{Dan1994} introduced the batching system, where a transmitter groups the users who request the same data in a short time and  transmits it together.  Another scheme, named patching, proposed in \cite{Hua1998}, allows users to join ongoing multicast transmissions at the time of their data request, patching missing pieces through additional allocated unicast transmissions. Furthermore, in \cite{Eager2001}, the authors proposed multicast stream merging for asynchronous data requests, assuming users have unlimited receiving bandwidth and buffering space. Additionally, in \cite{Lampiris2021}, a modified coded caching scheme was optimized to accommodate asynchronous data demands, demonstrating its ability to tolerate a degree of asynchronicity from a delay perspective. However, it is noteworthy that these investigations were carried out for wired networks without accounting for wireless environments.

The stochastic nature of wireless channels and the spatial distribution of users in wireless environments pose significant challenges for improving multicast transmission efficiency. Variations in wireless channel conditions, such as fading, noise, interference, and differences in user distances, can result in varying reception quality among users, making it challenging to ensure reliable and consistent reception quality, especially considering the diverse signal-to-noise-plus-interference ratios (SINRs) of users. The issue of multicasting in wireless environments has received considerable attention in existing literature. Some studies have explored the capacity of multi-input single-output (MISO) multicast channels under various conditions through asymptotic analysis \cite{Jindal2006, Park2010, Park2008}. In \cite{Sidiropoulos2006,Yue2016}, beamforming techniques for the MISO multicast channel to improve power efficiency were proposed. The authors in \cite{Kim2011,Wu2013} provided a transceiver beamforming algorithm to increase the transmission rate. Additionally, to address the challenge of single transmitter multicast to multiple groups of receivers, \cite{Karipidis2008, Mehanna2013, Xiang2014}  provided solutions, emphasizing optimal beamforming schemes. Addressing the challenge of acquiring accurate channel state information (CSI) at the transmitter side, some studies have investigated multicast schemes under imperfect statistical CSI at the transmitter (CSIT) \cite{Dai2015, Joudeh2015}. While these efforts have paved the way for implementing multicasting in wireless systems using beamforming or resource allocation techniques, they have primarily focused on synchronous  data requests.

Some studies explored multicast mechanisms in wireless communication, taking into account asynchronous data requests \cite{Song2018,Huang2012,Almowuena2016}. Authors in \cite{Song2018} presented a transmission scheme which aims at minimizing the total delivery time by adjusting unicast and multicast transmission rates when two users asynchronously request the same data.  In \cite{Huang2012}, bandwidth allocation between multicast transmission and unicast transmission is proposed to reduce the transmission power. Asynchronous multicast transmission based on batching scheme was investigated in \cite{Almowuena2016} to reduce energy consumption via resource allocation. However, these studies assume perfect CSIT but fail to provide guidelines on enhancing multicast transmission performance for asynchronous data requests in the absence of CSIT.
In order to ensure robustness across various scenarios, particularly when acquiring CSIT is challenging due to high mobility and low power of devices, the multicast mechanism should be designed to operate in the absence of CSIT \cite{Luo2020}.
While existing multicast mechanisms \cite{Dan1994,Hua1998,Eager2001} can be directly applied without CSIT, they were primarily proposed for wired channel conditions and thus do not address methods for reducing the average delivery time by considering the random requests of users and the variability of wireless channels.


In wireless network environments with asynchronous data requests, research on improving multicast transmission performance without using CSI has not yet been conducted, despite its potential to enhance multicasting across a wider range of scenarios.
To tackle these issues, this paper aims to design a novel wireless multicast mechanism to handle spatio-temporally asynchronous user requests in environments without CSIT.
More specifically, we propose a novel multicast mechanism comprising two phases. Firstly, the set-up phase allows us to harness the advantages of multicasting even in scenarios with asynchronous data requests. This phase entails collecting asynchronous file requests from users with different channel conditions and geographical locations, preparing for multicast transmission within a predetermined period referred to as the \textit{set-up time}. The users who request the same data during this set-up time are grouped together while receiving the requesting data in unicast mode, and then receive the data through multicast after the set-up time. By configuring different set-up times according to the popularity of the requested data, this mechanism effectively reduces unnecessary waiting times for users served by unicast transmission before entering into the multicast phase. 
Secondly, in the multicast phase, the remaining packets that were not delivered during the initial phase are transmitted in multicast mode, utilizing the aggregated bandwidth across users. This bandwidth aggregation allows us to achieve higher data transmission rates when using multicast, resulting in significantly improving transmission efficiency compared to conventional unicast transmission. To maximize the benefits derived from multicast transmission, we optimize the set-up time and transmission rates for both phases, considering wireless channel characteristics and the spatio-temporal randomness of user requests. Trade-off bounds between a packet size and outage probability is analyzed under random user distribution to optimize the transmission rate without CSIT. This optimization results in a novel multicast mechanism termed \textit{set-up based merged multicast (SMMC)}.
The contributions of this paper are summarized as follows.
\begin{itemize}
\item We introduce a new wireless multicast mechanism named SMMC, which involves grouping user requests during the set-up phase to enable multicast transmission for asynchronous data requests and merge allocated bandwidths for improving transmission efficiency. 
\item We establish both upper and lower bounds for the average delivery time by SMMC, taking into account the spatio-temporal randomness of user requests and wireless channel properties. Based on these bounds, we introduce an algorithm designed to optimize the set-up time and transmission rates for both unicast and multicast modes, aiming to minimize the expected file delivery time.
\item We introduce the fine-tuned SMMC mechanism to enhance SMMC further, utilizing posterior information regarding the number of users in the multicast group.
\item We conduct a comparative performance analysis of the proposed SMMC and fine-tuned SMMC in comparison to unicast transmission through mathematical analysis and numerical assessment. Moreover, we evaluate the effectiveness and feasibility of the proposed mechanisms while providing insights into the application of our SMMC mechanism to wireless networks.
\end{itemize}

The remainder of this paper is organized as follows. In Section II, we provide a comprehensive explanation of the system model under consideration, covering the network model, channel model, and transmission procedure of the SMMC mechanism. In Section III, we delve into the analysis of both the upper and lower bounds of the average SMMC delivery time and formulate the optimization problem. Section IV encompasses the analysis of the optimal transmission rate, the introduction of the algorithm to find the jointly optimized set-up time and transmission rates, and the presentation of fine-tuned SMMC to enhance the SMMC performance. Moving on to Section V, we numerically evaluate the performance of the SMMC and fine-tuned SMMC by delivery time metric. Finally, we conclude our work in Section VI.

\section{System Model \& SMMC Mechanism}
\subsection{Network and Channel Models }

\begin{figure}[!t]
\centering
\includegraphics[width=0.6\columnwidth]{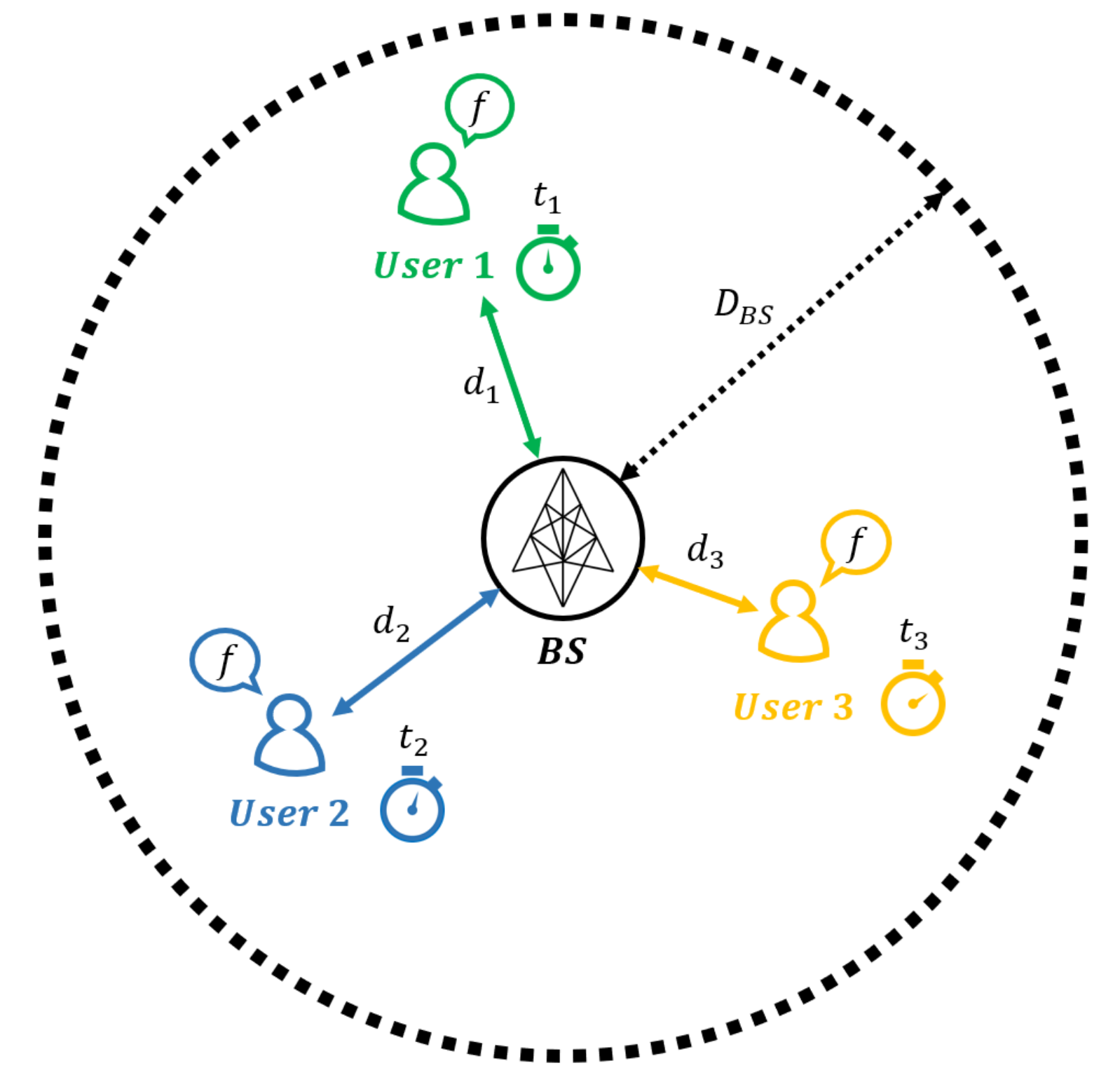}
\caption{Illustration of the system model with user requests for same file $f$ from multiple users ($K=3$). BS has the circular coverage region with radius $D_\text{BS}$ and request timeslot for each users ($t_1,t_2$ and $t_3$) are asynchronous.}
\label{Fig1}
\end{figure}
User scheduling typically takes place per base station (BS). For simplicity and without loss of generality, we concentrate on a wireless network featuring a single central BS located within a circular coverage area, illustrated in Fig. \ref{Fig1}. The radius of this coverage region is denoted as $D_\text{BS}$, with users uniformly distributed within this area. Each user generates a request for a file, denoted as $f$, with a data size of $L_f$ bits. These requests follow a Poisson arrival process characterized by an arrival rate of $\lambda_f$.
The BS transmits the requested file to the users via the wireless channel. The channel coefficient of user $k$ is denoted as $h_k$ and subject to Rayleigh fading since typical uniform scattering environments are under consideration, i.e. $h_k \sim \mathcal{CN}(0,1)$. The distance between user $k$ and BS is $d_k$. Then, the received signal-to-noise ratio (SNR) over the channel between the BS and user $k$, denoted as $\rho_k$, can be expressed as 
\begin{align}
    \rho_k &= \frac{|h_k|^2P_\text{BS}}{P_\text{noise}}\left({d_k}\right)^{-\eta}=\rho_0 |h_k|^2{d_k}^{-\eta}
\end{align}
where $\eta$ is the path-loss exponent, $P_\text{BS}$ is the transmit power and $P_\text{noise}$ is the noise power for allocated bandwidth $W$.
Then, the channel capacity is given by $W\log(1+\rho_k)$. An outage occurs if the predefined fixed transmission rate $R$ is larger than the capacity  \cite{Tse2005}.
Since the channel power gain, $|h_k|^2$, follows an exponential distribution, the outage probability for user $k$ at distance $d_k$ can be calculated as
\begin{align}
    \epsilon_k&=\textrm{Pr}[W\log(1+\rho_k)<R] \nonumber\\
    &=1-\exp\left(-\dfrac{2^{(\frac{R}{W})}-1}{\rho_0{d_k}^{-\eta}}\right)
\end{align}
From an operational perspective, in the event of an outage stemming from unfavorable channel conditions, retransmissions are performed using type-I hybrid automatic repeat request (H-ARQ) to manage unsuccessful receptions. 

The system operates on slot-based transmissions, where each time slot is of a duration $T_0$, mirroring the length of a packet, as a time slot encompasses a single packet. The channel is subject to block fading, where the channel gain remains constant during a slot but varies across slots. We assume perfect CSI at the receivers and the absence of CSIT.

\subsection{Set-up based Merged Multicast Mechansim}

\begin{figure}[!t]
\centering
\begin{subfigure}{\columnwidth}
  \centering
  \includegraphics[width=\textwidth]{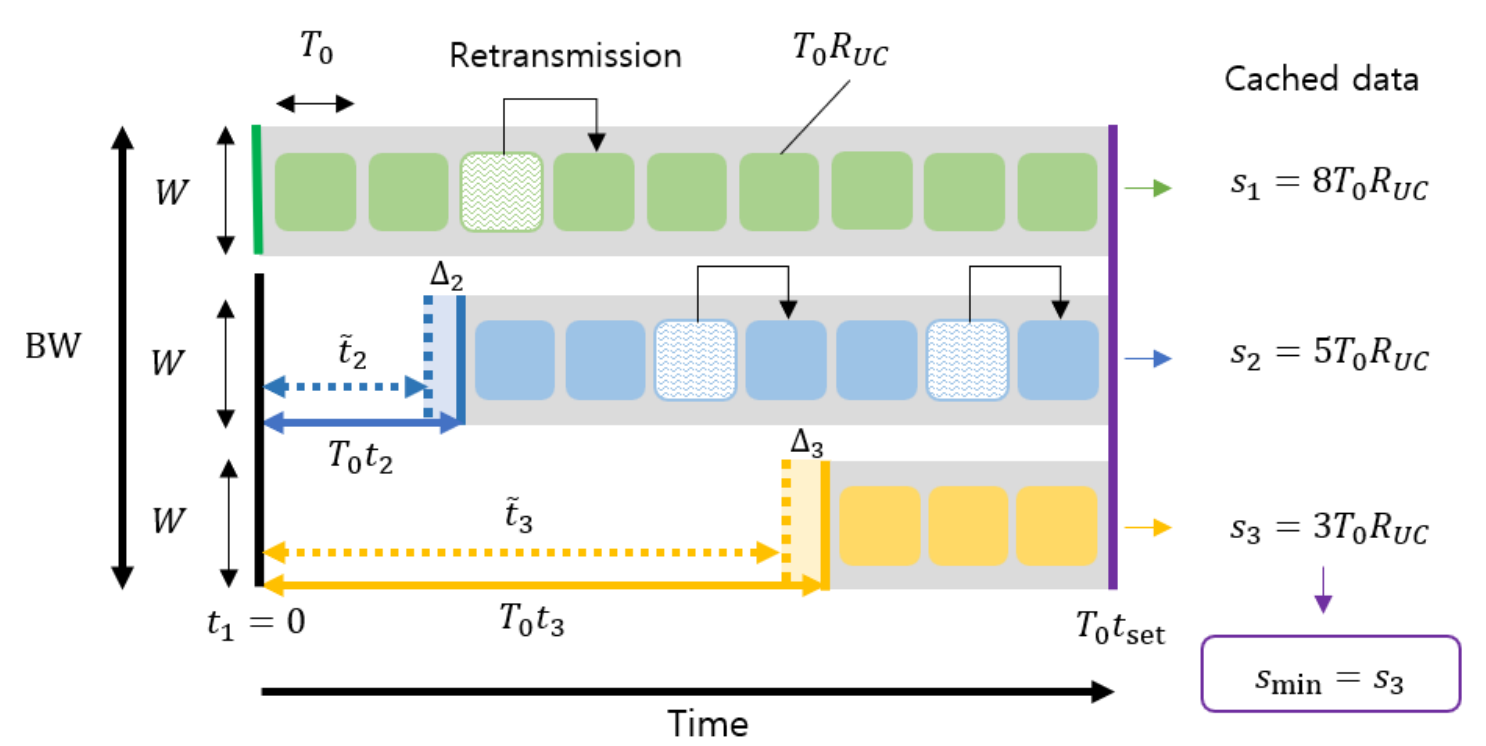}
  \caption{Set-up phase}
  \label{setup fig}
\end{subfigure}

\vspace{1pt}

\begin{subfigure}{\columnwidth}
  \centering
  \includegraphics[width=\textwidth]{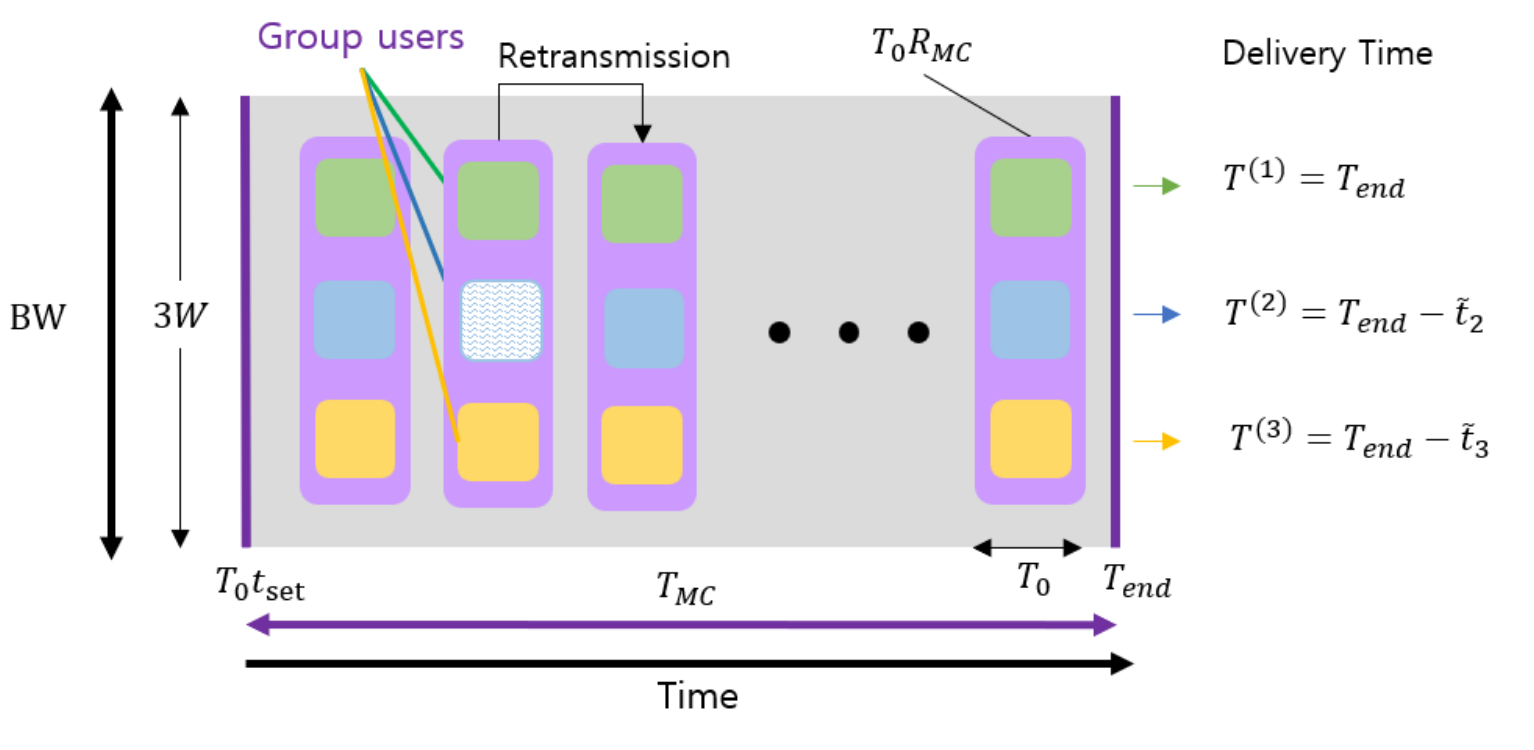}
  \caption{Multicast phase}
  \label{multicast fig}
\end{subfigure}
\caption{Illustration of the SMMC mechanism with (a) the set-up phase process and (b) the multicast phase process with $K=3$ users.}
\label{Fig2}
\end{figure}

Multiple users requesting the same file, denoted as $f$, prompt the implementation of multicasting as an efficient scheme for file transmission to each user. However, challenges arise due to the variability in request timing and outage probability among users, stemming from random request occurrences and diverse user channel conditions, including communication distances. Consequently, there is a pressing need for an efficient scheme to leverage multicasting in this asynchronous environment.

To address this challenge, we propose the SMMC mechanism, comprising a set-up phase and a multicast phase as illustrated in Fig. \ref{Fig2}. We introduce the concept of a set-up time, denoted as $t_{\mathrm{set}}$, representing the predetermined number of timeslots of the set-up phase. This phase commences when a user initiates a request for file $f$ from the BS. During the set-up phase, as in Fig.\ref{setup fig}, the BS aggregates additional requests for the same file from other users and sequentially transmits the file in packet units to each user via unicast, utilizing an assigned bandwidth $W$ and a predefined data rate $R_\text{UC}$.

The group of collected users is referred to as \emph{group users}, and a user index, denoted as $k \in \{1, 2, 3, \ldots, K\}$ is assigned based on the time order of their requests, and $K$ is the total number of group users. For example, in Fig.\ref{Fig2}, green, blue, and yellow users are indexed as 1, 2, and 3, respectively. If the elapsed time from the user 1's request to user $k$'s request is in the range $\tilde{t}_k \in (T_0t, T_0t+T_0]$ for $t\in \{0, 1, \ldots , t_{\mathrm{set}}-1\}$ and a single timeslot length $T_0$, the request timeslot of user $k$ is denoted as $t_k=t+1$ while $\tilde{t}_1 = t_1=0$. The gap between the exact requesting time and request timeslot is written as $\Delta_k = T_0t_k - \tilde{t}_k$. 
The packets successfully received by each user within the set-up time duration serve as cached data in the subsequent multicast phase, with the amount for user $k$ denoted by $s_k$. 
For example, the user 2 in Fig.\ref{setup fig} received 5 packets during the set-up phase, and $s_2=5T_0R_{\mathrm{UC}}$ since the size of a single packet is $T_0 R_{\mathrm{UC}}$.

Upon the completion of the set-up phase, the BS halts user collection and initiates the multicast phase. It begins transmitting file packets to the group users using multicast. To ensure the successful reception of the entire file by all users, multicast transmissions commence based on the least cached data among users, i.e., $s_{\min}$. During the multicast phase, the bandwidths of users in the group are merged, enabling packet transmission using the $KW$ bandwidth with data rate $R_\text{MC}$. Retransmissions persist until all group users successfully receive the packet. The whole process of the SMMC mechanism is summarized in \textbf{Algorithm \ref{algorithm2}} and on the following section, we will analyze the SMMC delivery time to optimize the transmission rates and the set-up time.





\section{SMMC delivery time Analysis}
Delivery time for successfully sending file $f$ to user $k$ can be defined as the elapsed time from the user request time $\tilde{t}_k$ until the file is successfully received by the user. Intuitively, the data rate influences delivery time by determining the number of packets needed to receive the entire file and the transmission outage probability. 
On the other hand, the set-up time $t_{\mathrm{set}}$ also affects the delivery time as it determines the number of user requests accommodated in the set-up phase. Hence, an optimal set-up time plays a crucial role in maximizing multicast efficiency. Conversely, an excessively prolonged set-up time results in inefficiency. With these considerations in mind, we derive the expected value of the average delivery time for SMMC across all group users. Additionally, we establish tractable convex bounds to assess the transmission rate and set-up time combination aimed at minimizing the expected delivery time.

\begin{algorithm}[!t]
\caption{SMMC}
\begin{algorithmic}[1]
    \State{$\textbf{Initialize: } L_f, \lambda_f, D_\text{BS}, \rho_0, W$,$R_\text{UC}$,$R_\text{MC}$,$t_{\mathrm{set}}$}
    \State{\texttt{|* Set-up Phase *|}}
    \State{BS receive a file $f$ request from a user}
    \State{$K=1$, $s_1=0$}
        \For {$t=1,2,...,t_{\mathrm{set}}$}
            \State{Unicast a packet to each user with rate $R_\text{UC}$}
            \For {$k=1,...,K$}
            \State{BS unicast a packet to user $k$ with rate $R_\text{UC}$}
            \State{\textbf{If} user $k$ receives a packet :}
            \State{\quad $s_k$ += $T_0 R_\text{UC}$}
            \EndFor
            \State{\textbf{If} Additional request for file $f$ arrive :}
            \State{\quad $n\gets$ number of additional requests for file $f$}
            \State{\quad $K = K +n$}
            \State{\quad $s_{K-n+1},...,s_{K}$ = $0$}
        \EndFor
        \State{$s_{\min}\gets\min\{s_1, ..., s_K\}$}
    \State{\texttt{|* Multicast Phase *|}}
        \State{Current cached data $\gets s_{\mathrm{min}}$}
        \State{\textbf{While} Current cached data  $\le$ $L_f$ \textbf{do}}
        \State{\quad \quad Multicast a packet to every users with rate $R_\text{MC}$}
        \State{\quad \quad \textbf{If} every users receive current packet \textbf{do}}
        \State{\quad \quad \quad Current cached data += $T_0 R_\text{MC}$}
\end{algorithmic}\label{algorithm2}
\end{algorithm}

\subsection{Problem Formulation}
Each user's delivery time is composed of the sum of the time spent in the set-up phase and multicast phase, as shown in Fig. \ref{Fig2}. Therefore, the delivery time for user $k$ is denoted as $T^{(k)}=T^{(k)}_\text{SU}+T^{(k)}_\text{MC}$ where $T^{(k)}_\text{SU}$ and $T^{(k)}_\text{MC}$ represent the time for the set-up and multicast phases, respectively. The time consumed for user $k$ during set-up phase can be expressed as 
\begin{align}
    T^{(k)}_\text{SU}=T_0(t_{\mathrm{set}}-t_k)+\Delta_k.
\end{align} 
On the other hand,  we can assume that $T^{(k)}_\text{MC}$ is the same for all the group users, i.e., $T^{(k)}_\text{MC}=T_\text{MC}$ $\forall k\in \{1,...,K\}$, because the delay difference caused by the last packet transmission is negligible compared to the overall file delivery time when the file data size is much larger than the data size in a single packet. Hence, we can derive the expected value of the  delivery time when the  number of group users is given by $K$: 
\begin{align}
    \mathbb{E}\left[T_{\mathrm{avg}}^K\right]&\!
=\mathbb{E}\left[\dfrac{\sum^K_{k=1}T^{(k)}_\text{SU}+T^{(k)}_\text{MC}}{K}\right] \nonumber\\
    &\!=\!\sum^K_{k=1}\dfrac{\mathbb{E}\left[T_0(t_{\mathrm{set}}-t_k)+\Delta_k\right]}{K}+\mathbb{E}\left[T_\text{MC}\right].\label{eq_uniform}
\end{align}  
To analyze this equation, we first present an important property of the Poisson process in the following lemma. 
 
\begin{lemma}\label{lemma0}
\cite{Taylor1998} Let $A_1, A_2 ,...$ be the occurrence times in a Poisson process of rate $\lambda>0$. Conditioned on total number of occurrence time until $t$ as $N(t)=K$, the random variables $A_1, A_2, \ldots A_K$ have the joint probability density  function
\begin{align}
    f_{A_1,...,A_K}(a_1,...,a_K | N(t) =K)=\dfrac{K!}{t^{K}} \nonumber \\
    for \quad 0<a_1 <\cdots < a_K \leq t. \nonumber
\end{align}
\end{lemma}
\begin{IEEEproof}
The proof is provided in Theorem 4.1 of \cite{Taylor1998}.
\end{IEEEproof}
This lemma means that the file request time instances of the given $K-1$ users during the set-up phase are uniformly distributed, $\sim\mathrm{Uniform} (0, T_0 t_{\text{set}})$. The timeslot at which a data request occurs 
 is also uniformly distributed in $\{1,2,...,t_{\mathrm{set}}\}$. Therefore, we can derive the first term of \eqref{eq_uniform} as follows.
\begin{align}
\sum^K_{k=1}&\dfrac{\mathbb{E}\left[T_0(t_{\mathrm{set}}-t_k)+\Delta_k\right]}{K} \nonumber\\
    &=\dfrac{T_0(t_{\mathrm{set}}K+t_{\mathrm{set}}-K+2)}{2K}
\end{align}  

The expected delivery time of multicast phase, $\mathbb{E}\left[T_\text{MC}\right]$, is affected by three main factors: the set-up time $t_{\mathrm{set}}$, the number of group users $K$, and the multicast transmission rate $R_\text{MC}$ for given file size $L_f$. Therefore, $\mathbb{E}[T_\text{MC}]$ can be expressed as a function of these individual components, denoted as $\tau(t_{\mathrm{set}}, K, R_\text{MC})$. Detailed analysis related to this function will be elaborated in Section \ref{section_Avg_MP}.

Since the file requests from users are temporally random, the number of group users $K$ is a random variable depending on $t_\text{set}$ and $\lambda_f$. By taking into account the Poisson probability distribution of $K$, we can express the expected value of the average  delivery time as
\begin{align}
T_{\mathrm{avg}}
&=\sum^{\infty}_{K=1}\dfrac{(t_{\mathrm{set}}\lambda_f)^{K-1}}{(K-1)!}\exp(-t_{\mathrm{set}}\lambda_f) \nonumber\\
    &\!\times\! \left(\!\dfrac{T_0(t_{\mathrm{set}}K+t_{\mathrm{set}}-K+2)}{2K}\!+\!\tau(t_{\mathrm{set}}, K, R_\text{MC})\!\right).
\end{align}

Consequently, we can formulate a problem for minimizing the average delivery time as follows:
\begin{align}
\mathbf{P1}: T^*_{\mathrm{avg}}=\min_{R_\text{UC},R_\text{MC},t_\text{set}}&T_\text{avg} \nonumber\\
\textrm{s.t.} \quad & 0 < R_\text{UC},R_\text{MC} \label{P1_rate}\\
&0\leq t_{\mathrm{set}} \leq L_f/(T_0R_\text{UC}) \label{P1_tset}
\end{align}
where \eqref{P1_rate} indicates the transmission rate constraint and constraint \eqref{P1_tset} denotes that maximum allowed set-up time is limited to the required time slots for a unicast transmission of the entire file without experiencing an outage. 

\subsection{Multicast Outage Probability with Distributed Users}\label{section_Avg_MP}
During the multicast phase, we utilize multicast transmission to send packets to the group users. If the user experiencing the minimum received SNR can successfully receive the packet, it ensures that all group users receive the packet without encountering an outage. Therefore, multicast capacity is determined by the link capacity of the worst receiver and we can obtain the outage probability during multicast for given $K$ users, denoted as $\epsilon_\text{MC}^K$, as follows
\begin{align}
    \epsilon_\text{MC}^K&=\textrm{Pr}\left[KW\log(1+\rho_{\mathrm{min}}^K)<R_\text{MC}\right]\label{error_mc}
\end{align}
where $\rho_{\mathrm{min}}^K$ represents the minimum received SNR among the $K$ group users. However, since the received SNR is influenced by both the user distance and the instantaneous channel gain, the exact value of $\epsilon_\text{MC}^K$ cannot be easily determined in a tractable form. Therefore, we resort to the following upper and lower bounds of the multicast outage probability to analyze the optimal transmission rate and set-up time.
\begin{prop}\label{proposition1}
The multicast outage probability with $K$ distributed users is upper bounded as
\begin{align}
    \epsilon_\text{\emph{MC}}^K\leq\hat{\epsilon}_\text{\emph{MC}}^K=1-\exp\left(-\dfrac{2^{(\frac{R_\text{\emph{MC}}}{KW})}-1}{\hat{\rho}_{\mathrm{min}}^K}\right)\label{Prop1}
\end{align}
where
\begin{align}
    \hat{\rho}_{\mathrm{min}}^K=\dfrac{\rho_0}{KD^{\eta}_\text{\emph{BS}}} \nonumber
\end{align}
\end{prop}
\begin{IEEEproof}
Since each channel power gain distribution is exponential, the minimum channel  power gain probability density function among $K$ group users, $f_K(x)$, can be written as 
\begin{align}
    f_K(x)&=\dfrac{dF_K(x)}{dx}=\dfrac{d}{dx}\left(1-(1-F_1(x))^K\right) \nonumber\\
    &=\dfrac{d}{dx}\left(1-\exp(-Kx)\right)=K\exp(-Kx).
\end{align}
where $F_K(x)$ is a cumulative distribution function of the minimum channel power gain among $K$ group users. 

On the other hand, the distance from the BS also affects the average received SNR. From this perspective, the worst-case scenario for the average received SNR occurs when the farthest user, positioned at a distance of $D_\text{BS}$, consistently experiences the minimum channel power gain among the $k$ users.  Then, we can get the lower bound of minimum received SNR ${\rho}_{\mathrm{min}}^K$ as follows:
\begin{align}
     {\rho}_{\text{min}}^K =  \min_{k\in \{1,...,K\}}\rho_0 d_k^{-\eta}|h_{k}|^2
     \geq \rho_0 D_\text{BS}^{-\eta} |h_{\mathrm{min}}|^2.\label{prop1_upperbound}
\end{align}
Consequently, we can derive the upper bound of the multicast outage probability, denoted as $\hat{\epsilon}_\text{MC}^K$, as follows:
\begin{align}
    \epsilon_\text{MC}^K
    &=\textrm{Pr}\left[\rho_{\mathrm{min}}^K<2^{(\frac{R_\text{MC}}{KW})}-1\right]\nonumber \\
    &\leq \textrm{Pr}\left[|h|_{\mathrm{min}}^2<\dfrac{2^{(\frac{R_\text{MC}}{KW})}-1}{\rho_0 D_\text{BS}^{-\eta}}\right] \label{prop1_proof1}\\ 
    &=1-\exp\left(-\dfrac{2^{(\frac{R_\text{MC}}{KW})}-1}{\hat{\rho}_{\mathrm{min}}^K}\right)=\hat{\epsilon}_\text{MC}^K.\label{prop1_proof2}
\end{align}
The inequality in \eqref{prop1_proof1} is derived from \eqref{prop1_upperbound} and the first equation in \eqref{prop1_proof2} holds since the minimum channel power gain $|h|_{\mathrm{min}}^2$ follows an exponential distribution.
\end{IEEEproof}

\begin{prop}\label{proposition2}
The multicast outage probability with $K$ distributed users can be lower bounded as
\begin{align}
    \epsilon_\text{\emph{MC}}^K&\geq\check{\epsilon}_\text{\emph{MC}}^K=1-\exp\left(-\dfrac{2^{(\frac{R_\text{\emph{MC}}}{KW})}-1}{\check{\rho}_{\mathrm{min}}^K}\right)\label{Prop2}
\end{align}
where,
\begin{align}
    \check{\rho}_{\mathrm{min}}^K=\dfrac{\rho_0(\eta+2)}{2KD_\text{\emph{BS}}^\eta}. \nonumber
\end{align}
\end{prop}
\begin{IEEEproof}
The multicast outage probability with $K$ distributed users is given by
\begin{align}
     \epsilon_\text{MC}^K
   &=\textrm{Pr}\left[\rho_{\mathrm{min}}^K<2^{(\frac{R_\text{MC}}{KW})}-1\right]\nonumber\\
    &=\!\int_{d_1}\!\cdots\!\int_{d_K}\!\underbrace{\left( \!1\!-\!\exp\left(\dfrac{-\left(2^{(\frac{R_\text{MC}}{KW})}-1\right)}{\left(\sum_k{(\rho_0 d_k^{-\eta})^{-1}}\right)^{-1}} \right)\!\right)}_{(a)} \nonumber\\
    &\quad \quad \quad \quad \quad \times p(d_1)\cdots p(d_K) \mathrm{d}d_1 \cdots \mathrm{d}d_K\label{prop2_equation}\\
    &=\mathbb{E}_{d_1,...,d_k}\left[1-\!\exp\left(\dfrac{-\left(2^{(\frac{R_\text{MC}}{KW})}-1\right)}{\left(\sum_k{(\rho_0 d_k^{-\eta})^{-1}}\right)^{-1}} \right)\right]\label{prop2_equation2}
\end{align}
where $p(d_{k})$ represents the probability density function of the distance between the BS and user $k$. By leveraging the property of exponential distributions for channel power gains, which states that the distribution of the minimum of exponential random variables with parameters $x_1, ..., x_K$ also follows an exponential distribution with a parameter $x_{\mathrm{min}}=x_1+\cdots + x_K$, we can represent the minimum channel  power gain distribution $\rho^K_{\mathrm{min}}$ for a given user distribution as term (a). We then derive  \eqref{prop2_equation} by integrating the term with respect to each user's distance distribution. Since the exponential function is convex, we can apply the Jensen's inequality to obtain the lower bound for \eqref{prop2_equation2} as follows.
\begin{align}
&\mathbb{E}_{d_1,...,d_k}\left[1-\!\exp\left(\dfrac{-\left(2^{(\frac{R_\text{MC}}{KW})}-1\right)}{\left(\sum_k{(\rho_0 d_k^{-\eta})^{-1}}\right)^{-1}} \right)\right]\nonumber\\
    &\geq 1-\exp\left(-\dfrac{\left(2^{(\frac{R_\text{MC}}{KW})}-1\right)\mathbb{E}\left[\sum_k{d_k^{\eta}}\right]}{\rho_0} \right)
\end{align}

As each user is uniformly distributed in a circular region with a radius of $D_\text{BS}$, we can easily derive the expected value of $\sum_k d_k^\eta$ as follows.
\begin{align}
\mathbb{E}\left[\sum_k{d_k^{\eta}}\right]&=K\int_0^{D_\text{BS}}d^{\eta} p(d) \mathrm{d}d  = \dfrac{2KD^{\eta}_\text{BS}}{\eta +2 }
\end{align}
Hence, we obtain the lower bound of multicast outage probability with $K$ distributed users as expressed in  \eqref{Prop2}.
\end{IEEEproof}

\subsection{Upper \& Lower Bounds for Average Delivery Time}
As discussed earlier, packet transmission begins from the subsequent packet following the minimum cached packet among the group of users during the setup phase, denoted as $s_{\min}$. If the minimum cached data size is small, the time to complete file transmission in the multicast phase will be longer due to a large remaining data amount, and vice versa. However, the cached data size for each user varies due to asynchronous file request times and dynamic channel gains between the BS and the users. In this subsection, we derive the lower and upper bounds of $\mathbb{E}[s_{\mathrm{min}}]$ for a given group of users and establish the upper and lower bounds of the average delivery time of SMMC based on these boundaries.

\begin{figure*}[h!]
\setcounter{equation}{21}
\small
\begin{align}\label{smin lowerbound}
\check{s}_{\mathrm{min}}\!=\!T_0 R_\mathrm{UC}\sum_{c=0}^{t_{\mathrm{set}}}\left(1\!-\!\sum_{c'=0}^c {t_{\mathrm{set}} \choose c'} {\epsilon^{(t_{\mathrm{set}}-c')}_{\max}}(1-\epsilon_{\max})^{c'}\right)\!\left(\dfrac{c-1+\sum_{t_k=0}^{t_{\mathrm{set}}-c}\sum_{c'=0}^{c} {t_\text{set}-t_{k} \choose c'} {\epsilon^{(t_{\mathrm{set}}-t_{k}-c')}_{\max}}(1-\epsilon_{\max})^{c'}}{t_{\mathrm{set}}} \right)^{K-1}
\end{align}
\normalsize
\hrulefill
\setcounter{equation}{20}
\end{figure*}

The upper bound of $\mathbb{E}[s_{\mathrm{min}}]$ can be derived in the following proposition, not as the minimum value among all $K$ users' cached data, but through an upper bound on the expected value of cached data from the last file request user.

\begin{prop}\label{proposition3}
The expected value of minimum cached data size for given group user number $K$ is upper bounded as
\begin{align}
    \mathbb{E}[&s_{\mathrm{min}}]\leq \hat{s}_{\mathrm{min}} \nonumber\\
    &=\dfrac{t_{\mathrm{set}}}{K} T_0 R_\text{\emph{UC}}\exp\left(\dfrac{-2(2^{(\frac{R_\text{\emph{UC}}}{W})}-1)D_\text{\emph{BS}}^{\eta}}{(\eta+2)\rho_0  
  }\right)\label{Prop3}
\end{align}
\end{prop}
\begin{IEEEproof}
The proof is provided in Appendix \ref{appendix1}
\end{IEEEproof}

The lower bound of $\mathbb{E}[s_{\mathrm{min}}]$ can be   derived in the following proposition by applying order statistics and considering the worst case scenario where every user is at the edge of the coverage region.
\begin{prop}\label{proposition4}
The expected value of minimum cached data size for given group user number $K$ is lower bounded as
\setcounter{equation}{22}
\begin{align}
    \mathbb{E}[s_{\mathrm{min}}]&\geq \check{s}_{\mathrm{min}}=\eqref{smin lowerbound}\label{Prop4} \nonumber
\end{align}

where $\epsilon_{\max}=1-\exp\left(-\dfrac{2^{(\frac{R_\text{\emph{UC}}}{W})}-1}{{\rho_0} D_\text{\emph{BS}}^{-\eta}}\right)$.
\end{prop}
\begin{IEEEproof}
The proof is provided in Appendix \ref{appendix2}
\end{IEEEproof}

However, the computation overhead of the lower bound  of $\mathbb{E}[s_{\mathrm{min}}]$ in \eqref{smin lowerbound} is not negligible. Therefore we use a simple assumption to simplify the lower bound computation.

\begin{assumption}\label{assumption1}
    If all users are equidistant from the BS, the user who requests data last will have the minimum cached data size.
\end{assumption}

Under this assumption, we can derive the lower bound  by considering $\mathbb{E}[s_{K}]$ when every user is at the edge of the coverage region. Consequently, the lower bound can be formulated in the following proposition. 

\begin{prop}\label{proposition5}
Based on Assumption \ref{assumption1}, the expected minimum cached data size for given group user number $K$ is lower bounded as
\begin{align}
    \mathbb{E}[&s_{\mathrm{min}}]\geq \check{s}_{\mathrm{min}} \nonumber\\
    &=\dfrac{t_{\mathrm{set}}-K}{K} T_0 R_\text{\emph{UC}}\exp\left(-\dfrac{2^{(\frac{R_\text{\emph{UC}}}{W})}-1}{{\rho_0} D_\mathrm{BS}^{-\eta}}\right)
\end{align}
\end{prop}
\begin{IEEEproof}
 The proof is provided in Appendix \ref{appendix3} 
\end{IEEEproof}

We will show that this assumption is acceptable by comparing the true minimum cached data size and the cached data size of the user whose request is the last among the group users  in Section \ref{Simulation}.

If the multicast outage probability for single packet transmission is given as $\epsilon^K_\text{MC}$, the expected  delivery time in the multicast phase is obtained as 
\begin{align}
\tau(t_{\mathrm{set}},\!K,\!R_\text{MC}\!)\!&=\! \sum_{i=0}^{t_\mathrm{set}} \!\textrm{Pr}\!\left[\dfrac{s_{\mathrm{min}}}{T_0 R_\text{UC}}\!=\!i\right]\!\cdot\!\dfrac{L_f-s_{\mathrm{min}}}{R_\text{MC}(1\!-\!\epsilon^K_\text{MC})}\nonumber\\
&=\dfrac{L_f-\mathbb{E}[s_{\mathrm{min}}]}{R_\text{MC}(1\!-\!\epsilon^K_\text{MC})}
\end{align}

Therefore, based on Propositions \ref{proposition1}-- \ref{proposition5}, we can establish both the upper and lower bounds for the expected delivery time in the multicast phase as follows:
\begin{align}
    \check{\tau}(t_{\mathrm{set}},\!K,\! R_\text{MC})\!\leq\!\tau(t_{\mathrm{set}},\!K,\! R_\text{MC})\!\leq\!\hat{\tau}(t_{\mathrm{set}},\!K,\!R_\text{MC})
\end{align}
where 
\begin{align}
    \check{\tau}(t_{\mathrm{set}}, K, R_\text{MC})=\dfrac{(L_f-\hat{s}_{\mathrm{min}})}{R_\text{MC}(1-\check{\epsilon}_\text{MC}^K)} \label{tau_lower}
\end{align}
and 
\begin{align}
    \hat{\tau}(t_{\mathrm{set}}, K, R_\text{MC})=\dfrac{(L_f-\check{s}_{\mathrm{min}})}{R_\text{MC}(1-\hat{\epsilon}_\text{MC}^K)}. \label{tau_upper}
\end{align}

Combining these results, the expected value of the  delivery time $T_{\mathrm{avg}}$  is upper bounded in a more tractable form as
\begin{align}
\hat{T}&_{\mathrm{avg}}=\sum^{\infty}_{K=1}\dfrac{(t_{\mathrm{set}}\lambda_f)^{K-1}}{(K-1)!}\exp(-t_{\mathrm{set}}\lambda_f) \nonumber\\
    &\!\times\! \left(\!\dfrac{T_0(t_{\mathrm{set}}K+t_{\mathrm{set}}-K+2)}{2K}\!+\!\dfrac{(L_f-\check{s}_{\mathrm{min}})}{R_\text{MC}(1-\hat{\epsilon}_\text{MC}^K)}\!\right)\label{hat_T}
\end{align}
Finally, we can reframe  the problem \textbf{P1} as a problem with the upper bounded but tractable objective function, as follows

\begin{align}
\mathbf{P2}: \hat{T}^*_{\mathrm{avg}}=\min_{R_\text{UC},R_\text{MC},t_{\mathrm{set}}}&\hat{T}_{\mathrm{avg}} \nonumber\\
\textrm{s.t.} \quad & 0 < R_\text{UC},R_\text{MC} \nonumber \\
&0\leq t_{\mathrm{set}} \leq L_f/T_0R_\text{UC} \nonumber 
\end{align}

Although it is also possible to recast the problem \textbf{P1} with a lower bounded objective function, we will concentrate on analyzing \textbf{P2} in Section \ref{optimize section} since the upper bound offers more valuable insights regarding the delivery time.


\section{Optimal Transmission Rates $\&$ Set-up time}\label{optimize section}
In this section, we optimize the transmission rates for both the set-up phase and the multicast phase, as well as the set-up time to minimize the upper-bounded average delivery time of SMMC. We begin by proving the convexity of the objective function of the problem \textbf{P2}  with respect to transmission rates. Subsequently, we present an algorithm to determine the jointly optimal values of the transmission rates and set-up time. After that, we introduce a fine-tuned SMMC approach which leverages information about the number of users within a group after the set-up phase to achieve further performance enhancements.

\subsection{Optimal transmission rate analysis}
To demonstrate the convexity of the objective function in the problem $\mathbf{P2}$ with respect to $R_\text{UC}$ and $R_\text{MC}$, our first step involves establishing properties of the double exponential function in Lemmas \ref{lem1} and \ref{lem2}.

\begin{lemma}
A general form of double exponential function given by  $f(x)=\exp \left(\dfrac{b^x}{a}\right)$ satisfies 
\label{lem1}
\begin{align}
f(x)> 1, \quad f'(x)&> 0, \quad f''(x)> 0, \label{lem1_e1}\\ 
\dfrac{f'(x)}{f(x)}&< \dfrac{f''(x)}{f'(x)}. \label{lem1_e2}
\end{align}
for $x\ge0$ when $a (>0)$ and $b (>1)$ are constants. 
Here, $f'(x)=df(x)/dx$ and $f''(x)=d^2f(x)/dx^2$ denote the first and second order derivatives of $f(x)$, respectively.
\end{lemma}

\begin{IEEEproof}
The proof of \eqref{lem1_e1} is straightforward, so our focus is on demonstrating \eqref{lem1_e2}. We start by calculating $f'(x)$ and $f''(x)$ as follows
\begin{align}
f'(x)&=\dfrac{\ln b}{a} b^x \exp\left(\dfrac{b^x}{a}\right), \label{lem1_p1}\\ 
f''(x)&=\dfrac{\ln b}{a}b^x\exp \left(\dfrac{b^x}{a}\right)\left(\ln b + \dfrac{\ln b}{a}b^x \right) \label{lem1_p2}
\end{align} 
This leads to the expressions
\begin{align}
\dfrac{f'(x)}{f(x)}&=\dfrac{\ln b}{a}\cdot b^x, \label{lem1_p3}\\ 
\dfrac{f''(x)}{f'(x)}&=\ln b + \dfrac{\ln b}{a} \cdot b^x. \label{lem1_p4}
\end{align} 
Since $a>0$ and $b>1$, \eqref{lem1_e2} holds for $x\ge0$. 
\end{IEEEproof}

\begin{lemma}
For the double exponential function given by $f(x)=\exp \left(\dfrac{b^x}{a}\right)$, where $a (>0)$ and $b (>1)$ are constants, the function $\dfrac{f(x)}{x}$ is convex for $x>0$.
\label{lem2}
\end{lemma}

 \begin{IEEEproof} To demonstrate the convexity of $\dfrac{f(x)}{x}$, we first calculate its second derivative as follows.
\begin{align} 
\dfrac{d^2}{dx^2}\left(\dfrac{f(x)}{x}\right)=\dfrac{2f(x)}{x^3}-\dfrac{2f'(x)}{x^2}+\dfrac{f''(x)}{x}.\label{lem2_p1}
\end{align} 
According to Lemma \ref{lem1}, the feasible set of $f'(x)$ is $0<f'(x)$. Therefore, the proof is established by demonstrating that this lemma holds true for the following three cases.

\begin{enumerate} 
\item[(i)] Case 1 $\left( 0<f'(x)<\dfrac{f(x)}{x} \right)$ :
\begin{align}
\dfrac{d^2}{dx^2}\left(\dfrac{f(x)}{x}\right)=\dfrac{2}{x^2}\left(\dfrac{f(x)}{x}-f'(x)\right)+\dfrac{f''(x)}{x} \label{lem2_p2}
\end{align} 
The first term on the right-hand side of \eqref{lem2_p2} is positive, and $f''(x)> 0$ from Lemma 1. 

\item[(ii)]  Case 2 $\left(\dfrac{f(x)}{x}\le f'(x)<2 \cdot \dfrac{f(x)}{x} \right)$:
\begin{align}
\dfrac{d^2}{dx^2}\left(\dfrac{f(x)}{x}\right)&=\dfrac{1}{x^2}\left(\dfrac{2f(x)}{x}-f'(x)\right)+\dfrac{f''(x)}{x}-\dfrac{f'(x)}{x^2} \label{lem2_p3} \\
&\stackrel{(a)}{\ge} \dfrac{1}{x}\left(f''(x)-\dfrac{f'(x)}{x}\right) \stackrel{(b)}{>} 0\label{lem2_p4}
\end{align} 
The first inequality (a) in \eqref{lem2_p4} holds since the first term on the right-hand side of equality in \eqref{lem2_p3} is positive, and the second inequality (b) holds based on \eqref{lem1_e2} in Lemma 1 for the range condition of this case. 

\item[(iii)]  Case 3 $\left( 2 \cdot \dfrac{f(x)}{x}\le f'(x) \right)$: 
\begin{align}
\dfrac{d^2}{dx^2}\left(\dfrac{f(x)}{x}\right)& \ge \dfrac{f''(x)}{x}-\dfrac{2f'(x)}{x^2}\label{lem2_p5}\\
&=\dfrac{1}{x}\left(f''(x)-\dfrac{2f'(x)}{x}\right) > 0  \label{lem2_p6}
\end{align} 
In this case, \eqref{lem2_p5} is due to the property in \eqref{lem1_e1} of Lemma 1 and \eqref{lem2_p6} follows from the range condition in this case. 
\end{enumerate}
Combining all three cases above, we can prove that the second derivative of $f(x)/x$ is always positive for $x>0$.
\end{IEEEproof}

Next, we intend to optimize the unicast transmission rate in the set-up phase, aiming to minimize the upper-bounded average delivery time using the two previous lemmas. The upper-bounded average delivery time is a decreasing function of the minimum cached data size, $s_{\mathrm{min}}$. This means that the optimal value of the unicast transmission rate $R_\text{UC}$ is the value that maximizes the value of $s_{\mathrm{min}}$. According to Proposition \ref{proposition5}, we can rewrite the lower bound of the expected minimum cached data size, which is incorporated in the upper-bounded average delivery time,  as
\begin{align}
    \check{s}_{\mathrm{min}}=\frac{t_{\mathrm{set}}T_0}{K}\left[\frac{1}{R_\text{UC}}\exp\left(\dfrac{2^{(\frac{R_\text{UC}}{W})}-1}{\rho_0 D_\text{BS}^{-\eta}}\right)\right]^{-1}.\label{smin_inv}
\end{align}
Then we can formulate the optimization problem to find $R_\text{UC}^*$ as 
\begin{align}
    R_\text{UC}^*=\argmin_{R_\text{UC}}&\frac{1}{R_\text{UC}}\exp\left(\dfrac{2^{(\frac{R_\text{UC}}{W})}-1}{\rho_0 D_\text{BS}^{-\eta}}\right)\label{UC_opt}\\
    \textrm{s.t.} \quad & 0 < R_\text{UC}.  \nonumber
\end{align}
The objective function of this problem is expressed in the form $f(R_\text{UC})/R_\text{UC}$ where $f(x)$ is a double exponential function to $x$. Therefore, based on Lemma \ref{lem2}, the problem is convex since the constraint and the objective function are both convex for $R_\text{UC}$. Therefore,  it can be efficiently solved by using the convex solvers such as CVX \cite{Grant2014}. 

On the other hand, the multicast transmission rate also affects the upper-bounded average delivery time. Since \eqref{hat_T} is an increasing function with respect to \eqref{tau_upper} and the time for set-up phase  is independent of the multicast transmission rate,  we can set up the optimization problem to find $R_\text{MC}^*$ which minimizes the value of \eqref{hat_T} as follows.
\begin{align}
R_\text{MC}^*=\argmin_{R_\text{MC}}&\sum^{\infty}_{K=1}\dfrac{(t_{\mathrm{set}}\lambda_f)^{K-1}}{(K-1)!}\exp(-t_{\mathrm{set}}\lambda_f)\nonumber\\
    &\times\left(\dfrac{L_f-\check{s}_{\mathrm{min}}}{R_\text{MC}}\right)\exp\left(\dfrac{2^{(\frac{R_\text{MC}}{KW})}-1}{\hat{\rho}_{\mathrm{min}}^K}\right).\label{MC_opt}\\
    \textrm{s.t.} \quad & 0 < R_\text{MC}. \nonumber
\end{align}
Note that the objective function of this problem is a linear summation of the functions in the form of $f(R_\text{MC})/R_\text{MC}$ where $f(x)$ is a double exponential function for $x$. Therefore, this problem is a convex problem based on Lemma \ref{lem2} and can also be solved by an existing convex solver. However, because the range of $K$ is infinite, we have to limit the range of $K$ when we actually solve the problem. Nevertheless, we can select an appropriate maximum value of the number of group users as $K_{\mathrm{max}}$, for given $t_{\mathrm{set}}\lambda_f$, such that the probability that the number of group users becomes larger than $K_{\mathrm{max}}$ remains less than an arbitrary small value $\delta \approx 0$.

\begin{algorithm}[!t] 
\caption{Searching Optimal $R_\text{UC}$, $R_\text{MC}$ and $t_\text{set}$}
\begin{algorithmic}[1]
    \State{$\textbf{Initialize: } L_f, \lambda_f, D_\text{BS}, \rho_0, W$}
    \State{$\hat{T}_{\mathrm{avg}}^*=\infty$}
    \State{$R_\text{UC}^*\gets$Pre-fixed unicast data rate based on \eqref{UC_opt}.}
    \State{$t_{\max}=\lceil L_f/R_\text{UC}^* \rceil$}
        \For {$t=0,1,...,t_{\max}$}
            \State{$t_{\mathrm{set}}\gets t$}
            \State{$R_\text{MC}^t\gets$optimal transmission rate based on \eqref{MC_opt}.}
            \State{$\hat{T}_{\mathrm{avg}}\gets$\eqref{hat_T} for given $R_\text{MC}^t$}
            \State{$\textbf{If }$ $\hat{T}_{\mathrm{avg}}\le \hat{T}_{\mathrm{avg}}^*$ \textbf{do}}
            \State{ \quad $\hat{T}_{\mathrm{avg}}^*\gets \hat{T}_{\mathrm{avg}}$}
            \State{ \quad $R_\text{MC}^*\gets R^t_\text{MC}$, $t_{\mathrm{set}}^*\gets t_{\mathrm{set}}$}
        \EndFor
    \State{\textbf{Return} $R_\text{UC}^*$,$R_\text{MC}^*$,$t_{\mathrm{set}}^*$}
\end{algorithmic}\label{algorithm1}
\end{algorithm}

\subsection{Set-up time selection and fine-tuned SMMC}

To minimize the upper bounded average delivery time by jointly considering the transmission rates and the set-up time, we introduce an algorithm to find the optimal solution of \textbf{P2} in \textbf{Algorithm 2}. 
The BS determines the optimal unicast transmission rate based on problem \eqref{UC_opt} (line 3). The optimal unicast transmission rate in \eqref{UC_opt} remains unaffected by the specific file being requested, regardless of the file's size $L_f$ or popularity $\lambda_f$. This allows the base station to apply a consistent unicast transmission rate for all file requests. As a result, the BS only needs to calculate $R_\text{UC}^*$ once for the given $\rho_0, W, \eta$ and $D_\text{BS}$.

After obtaining $R_\text{UC}^*$, we set the maximum value of set-up time (line 4). This is the minimum number of time slots required to successfully receive the file only through unicast. The maximum value for the set-up time can be adjusted according to specific environmental constraints. Subsequently, for each set-up time value $t_{\mathrm{set}}=t$, the BS calculates the optimal multicast transmission rate $R_\text{MC}^t$ by solving \eqref{MC_opt} (lines 7). After that, the BS acquires the upper bound on average delivery time for spatio-temporally distributed random user requests by substituting $t_{\mathrm{set}}$ and $R_\text{MC}^t$ in  \eqref{hat_T}  (lines 8). During the iterative process, BS updates the value of $R_\text{MC}^*$ and $t_{\mathrm{set}}^*$ if the delivery time calculated in the current iteration is the minimum value encountered so far (lines 9-11). Finally, after all iterations have concluded, we obtain the optimal values (line 13). Since the time slot is a discrete variable, the search space that the BS needs to explore is finite. Furthermore, The process of finding the optimal transmission rate and set-up time using this algorithm can be performed offline. Consequently, this computational process does not introduce any additional delays in transmitting files to users in real-time scenarios.

It's important to note that the optimal solution of \textbf{P2} is calculated in a statistical way based on the spatial-temporal distribution of the random user requests.  However, once the set-up phase is completed, the BS can harness the information about the exact number of group users. This enables the BS to fine-tune the multicast transmission rate for the exact number of group users. This adjustment method is referred to as \textit{fine-tuned SMMC}, and the  multicast transmission rate  is re-optimized as
\begin{align}
    R_\text{MC}^{**}=\argmin_{R_\text{MC}} \left(\dfrac{1}{R_\text{MC}}\right)\exp\left(\dfrac{2^{(\frac{R_\text{MC}}{KW})}-1}{\hat{\rho}_{\mathrm{min}}^K}\right)\label{MC_refine}
\end{align}
Based on Lemma \ref{lem2}, \eqref{MC_refine} is also a convex function with respect to $R_\text{MC}$ and consequently, $R_\text{MC}^{**}$ can be obtained by using an existing convex solver. 
In the fine-tuned SMMC, the BS re-optimizes the multicast transmission rate based on \eqref{MC_refine} after the set-up phase (line 10 in Algorithm 1) ends. Through this process, we can enhance multicast performance, requiring only a simple computation before the multicast phase begins.

\section{Simulation Results}\label{Simulation}

\begin{figure}[t]
\centering
	\includegraphics[width = 0.9\columnwidth]{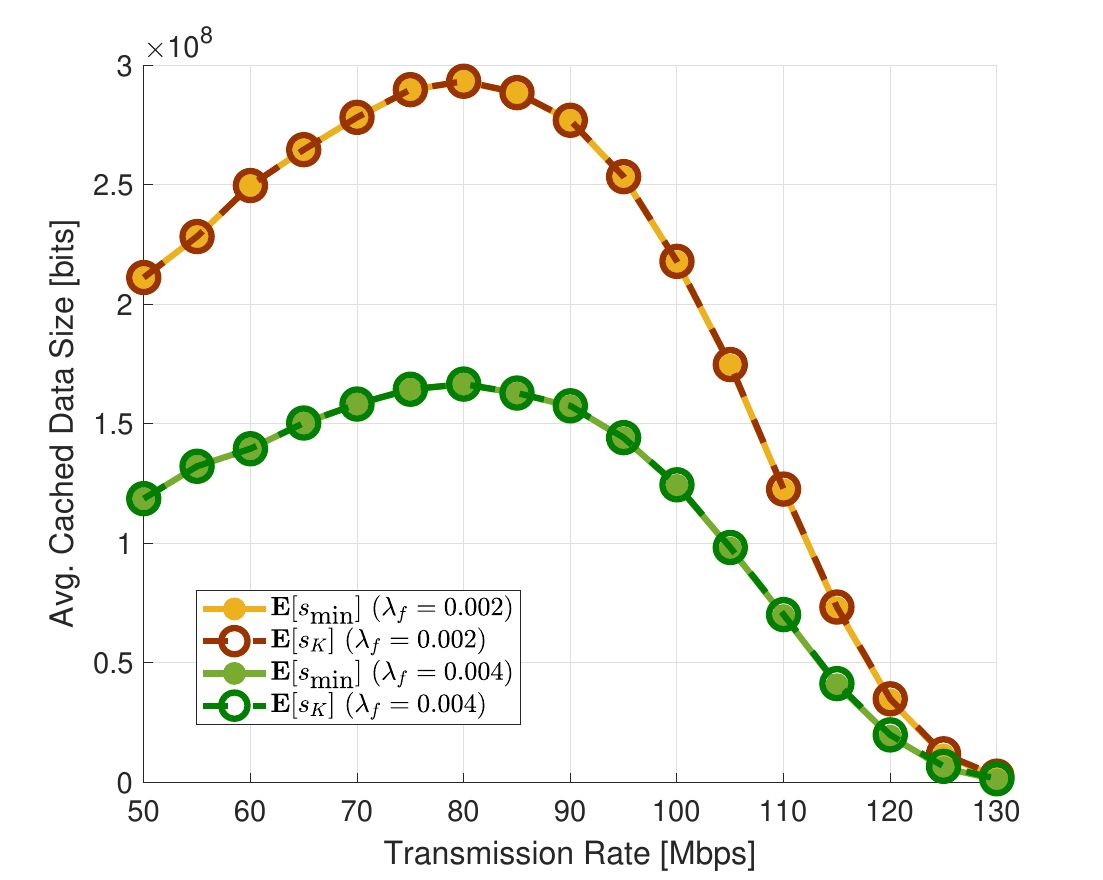}
	\caption{The averaged minimum cached data size in the set-up phase  (solid line) and average cached data size of the last requesting user (dotted line) for different transmission rates ($R_\text{UC}$) at $\lambda_f = 0.002$ and $\lambda_f = 0.004$. }
	\label{fig:Sim1}
\end{figure}

In this section, we conduct a numerical performance evaluation of the proposed SMMC and fine-tuned SMMC, validating the analytical results presented in the preceding sections. We assume that the BS allocates a equal bandwidth of $W = 10$ MHz and equal transmit power $P_{\mathrm{BS}} = 500$ mW  to each requesting user while the noise power is $P_{\mathrm{noise}} = -104$dBm. Additionally, we consider a fixed path loss exponent, $\eta = 4$, and a time slot length of $T_0 = 10$ milliseconds. The coverage radius of the BS is defined as $D_\text{BS} = 300$ meters, and the target data size is $L_f = 1$ Gigabytes.

\begin{figure}[!t]
\centering
\begin{subfigure}{0.9\columnwidth}
  \centering
  \includegraphics[width=\textwidth]{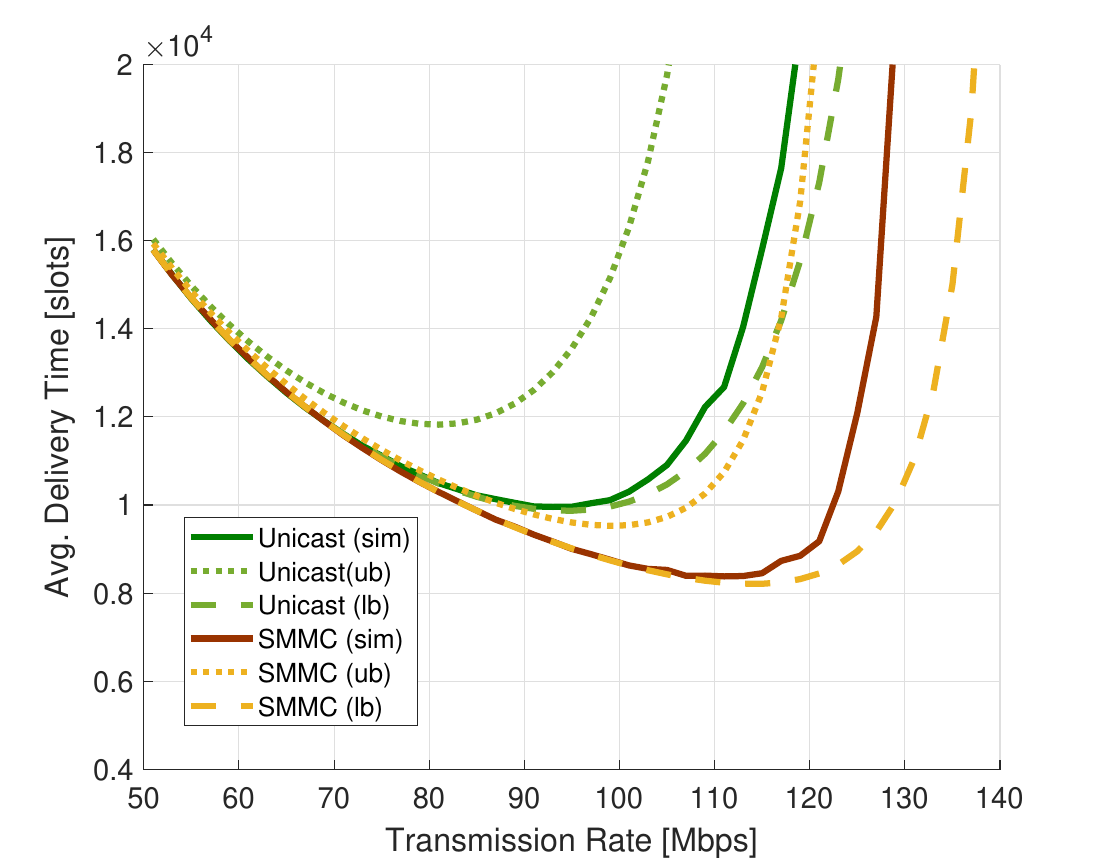}
  \caption{}
  \label{fig:Sim2a}
\end{subfigure}
\begin{subfigure}{0.9\columnwidth}
  \centering
  \includegraphics[width=\textwidth]{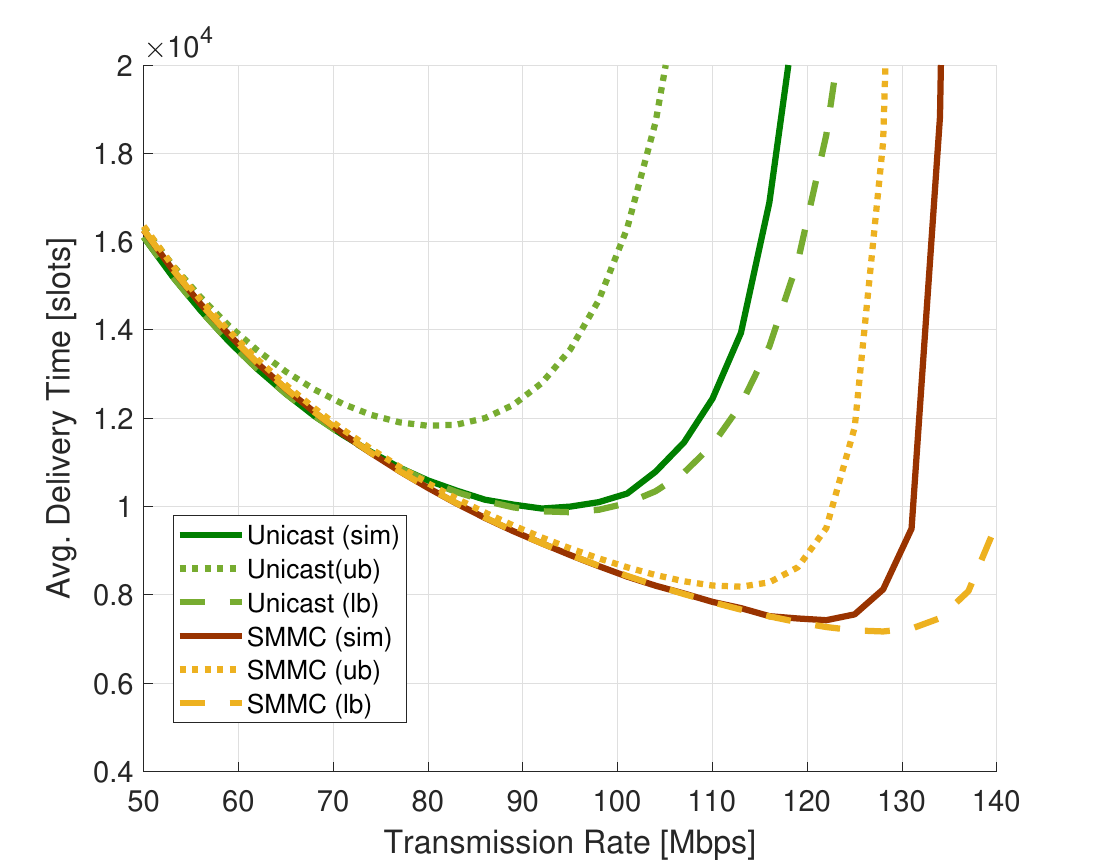}
  \caption{}
  \label{fig:Sim2b}
\end{subfigure}
\caption{Comparison of the upper bound (ub), lower bound (lb) and iterative simulation results (sim) of unicast and SMMC average delivery time versus transmission rate ($R_\text{MC}$) when user request frequency is (a) $\lambda_f=0.002$ and (b) $\lambda_f=0.004$  with a given set-up time $t_{\mathrm{set}}=1000$.}
\label{fig:Sim2}
\end{figure}

First, we compare the average minimum cached data size in Fig. \ref{fig:Sim1} to demonstrate the validity of our Assumption 1. Based on the assumption, all users are equidistant from the BS with $d_k = 300$ m for $\forall {k\in [1, K]}$. Users request a file asynchronously during $t_{\mathrm{set}}=1,000$ time slots. We plot the average number of cached packets in the set-up phase by running 1,000 realizations for two request frequency cases. In particular, the solid line shows the minimum cached data size among $K$ users and the dashed line exhibits the cached data size of the last requesting user. From the figure, we can see that the average sizes of the last user's cached data closely approximates the average minimum cached data sizes. Hence, we can argue that  Proposition \ref{proposition5} based on Assumption \ref{assumption1} will closely align with the actual value.
Furthermore, this figure allows us to observe the relationship between the unicast transmission rate and the minimum cached data size. If the transmission rate is lower than the optimal rate, even though the outage probability decreases, the reduction in the size of a single packet has a more dominant effect. Therefore, the average minimum cached data size decreases linearly. Conversely, if the transmission rate becomes higher than the optimal rate, the outage probability increases, resulting in fewer successful packet receptions. Consequently, the cached data size approaches to zero.

\begin{figure}[!t]
\centering
\begin{subfigure}{0.9\columnwidth}
  \centering
  \includegraphics[width=\textwidth]{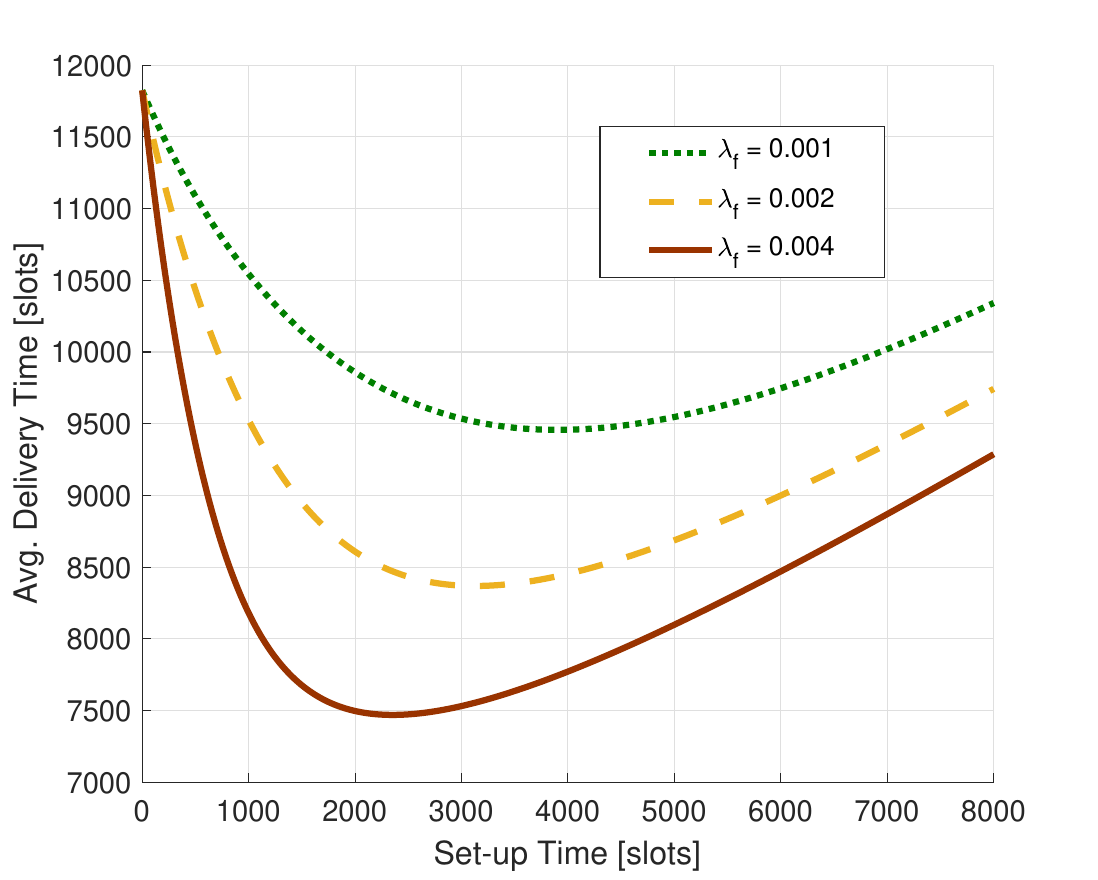}
  \caption{}
  \label{fig:Sim3a}
\end{subfigure}
\begin{subfigure}{0.9\columnwidth}
  \centering
  \includegraphics[width=\textwidth]{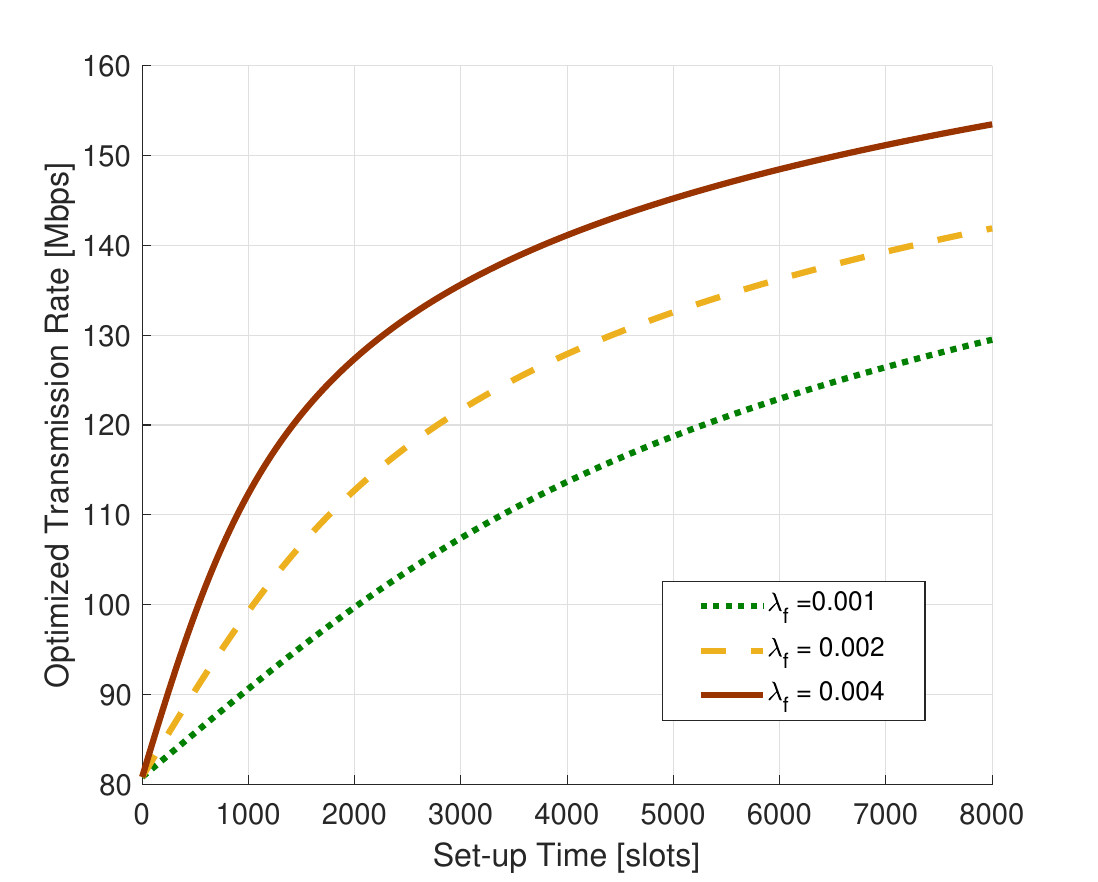}
  \caption{}
  \label{fig:Sim3b}
\end{subfigure}
\caption{(a) The analyzed average delivery time and (b) optimized transmission rate ($R_\text{MC}$) versus the set-up time ($t_\text{set}$) for different user request frequencies $\lambda_f=0.001, 0.002, 0.004$.}
\label{fig:Sim3}
\end{figure}

Fig. \ref{fig:Sim2} depicts the analyzed upper bound (ub), lower bound (lb), and the simulation result (sim) for the average delivery time in unicast and SMMC scenarios, with respect to the multicast phase transmission rate, $R_\text{MC}$ under a fixed set-up time of $t_{\mathrm{set}}=1,000$. We consider the average delivery time for unicast as a special case of SMMC with a single group user without set-up phase ($t_{\mathrm{set}}=0$). As observed, the simulation results lie within the range defined by analyzed upper and lower bounds. The minimum average delivery time achieved through SMMC is nearly $20\%$ lower than that of unicast, since the delivery time can be significantly reduced when the request frequency is higher. Furthermore, by comparing Fig.\ref{fig:Sim2a} and Fig.\ref{fig:Sim2b}, it is evident that the minimum average delivery time decreases as user request frequency increases. This is because higher request frequencies enable the formation of larger user groups, which, despite reducing the minimum channel gain due to the increase in the number of users, can still achieve higher transmission rates by utilizing a wider bandwidth.

The analyzed upper bound for the average delivery time of the SMMC and the optimized transmission rate for set-up time are respectively illustrated in Fig. \ref{fig:Sim3}. In Fig. \ref{fig:Sim3a}, it is evident that the average delivery time decreases as $t_{\mathrm{set}}$ increases up to a certain point, beyond which it starts to rise. This is because a longer set-up phase allows more opportunities to gather group users and leverage the advantages of multicast. However, an excessively long set-up phase ultimately delays the start of the multicast phase, resulting in increased delivery time. The optimal value of $t_{\mathrm{set}}$ is 3,901, 3,128, 2,353 slots for $\lambda_f = 0.001, 0.002, 0.004$, respectively. This implies that if the requested data is more popular, a shorter set-up phase is sufficient to harness multicast benefits, and the delivery time improvement with SMMC also grows as the request frequency increases. According to  Fig. \ref{fig:Sim3b}, the optimal transmission rate increases with higher values of set-up time and request frequency because the expected number of group users also increases in these cases.

\begin{figure}[t]
\centering
	\includegraphics[width = 0.9\columnwidth]{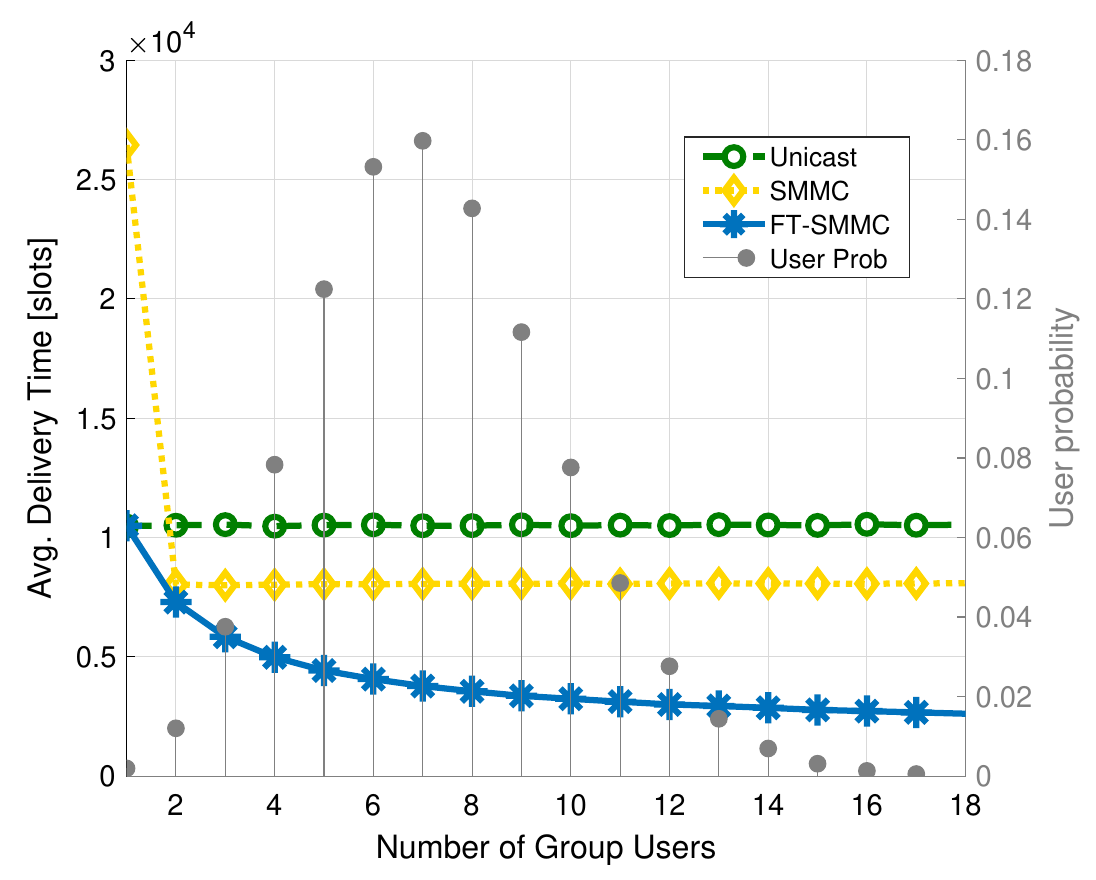}
	\caption{The average delivery time for unicast, SMMC, fine-tuned SMMC (FT-SMMC) and probability distribution for each number of group users under optimized set-up time ($t_{\mathrm{set}}$) and transmission rate ($R_\text{MC}$) when $\lambda_f = 0.002$. }
	\label{fig:Sim4}
\end{figure}

Fig. \ref{fig:Sim4} illustrates the simulated average delivery time versus the number of group users  for unicast, SMMC, and fine-tuned SMMC (FT-SMMC), along with the probability distribution of the corresponding  number of users for a given request frequency, $\lambda_f = 0.002$. The set-up time is fixed as the optimized value of $t_{\mathrm{set}}=3128$ slots, and we adopt $R_\text{MC}^*=122.6$ Mbps for the SMMC derived by equation \eqref{MC_opt}. In unicast, the BS transmits to each user by unicast transmission with $R_\text{UC}^*$ transmission rate. The results demonstrate that the SMMC mechanism consistently reduces average delivery time by approximately $20\%$ compared to unicast when there are two or more group users. However, when there is only a single user in a group, unicast achieves lower delivery time. This is because, in the SMMC, the transmission rate is optimized for multiple group users based on the probability distribution of the number of group users, which depends on $\lambda_f$ and $t_{\mathrm{set}}$, while in unicast, the transmission rate is optimized for a single user. However, the probability of a single user group is quite low, around $0.19\%$ in this case, and in a long-term perspective, the SMMC can provide better delivery time performance for all users. 
Moreover, it demonstrates that optimizing the transmission rate after the set-up phase ends for the specific number of users as in our proposed fine-tuned SMMC method leads to the lowest average delivery time across all group user number scenarios.

\section{Conclusion}
We have introduced and analyzed an efficient multicast mechanism tailored for addressing asynchronous data requests. Our proposed mechanism, termed SMMC, leverages a set-up based approach to efficiently manage the temporal variability in user requests, allowing for the aggregation of the allocated bandwidths.  To comprehensively assess system performance, we've derived upper and lower bounds for average delivery time, accounting for spatiotemporal randomness in user requests and wireless channel characteristics. Furthermore, we have extended the SMMC approach by incorporating information about the number of users within a group, resulting in a mechanism we refer to as fine-tuned SMMC.  Our simulation results demonstrate that in no CSIT wireless communication scenarios with multiple users requesting the same data, the SMMC technique offers a considerable reduction in average delivery time compared to traditional unicast-based data transmission, effectively addressing the challenges posed by spatiotemporal randomness in user requests.

%

\appendices
\def\thesection{\Alph{section}}%
\def\thesectiondis{\Alph{section}}%

\section{Proof of Proposition \ref{proposition3}} \label{appendix1}
\setcounter{equation}{0}
\renewcommand{\theequation}{A.\arabic{equation}}
By the definition of the minimum cached data size among group users and applying the lower bound of outage probability, we can derive the upper bound of $\mathbb{E}[{s_{\mathrm{min}}}]$ as follows.
\begin{align}
    \mathbb{E}[{s_{\mathrm{min}}}] &= \mathbb{E}[\text{min}\{s_1,..., s_K \}]\leq \mathbb{E}[s_K]\\
    =\int_{d_K}\!&\!\sum_{t=0}^{t_{\mathrm{set}}-1}\!\Pr\left[t_K\!=\!t\right](t_{\mathrm{set}}\!-\!t)(1\!-\!\epsilon_K)T_0 R_\text{UC} \mathrm{d}d_{K} \\
    &=\int_{d_K}(1-\epsilon_K)T_0R_\text{UC} \left[t_{\mathrm{set}} - \mathbb{E}[t_K] \right]\mathrm{d}d_{K}
\end{align}
Based on Lemma \ref{lemma0}, we can derive the expectation of the last timeslot among the requests from $K$ users as
\begin{align}
    \mathbb{E}[t_K]&=\sum_{t=1}^{t_{\mathrm{set}}}  \Pr\left[t_K \geq t\right] =\sum_{t=1}^{t_{\mathrm{set}}} 1-\left(\dfrac{t-1}{t_{\mathrm{set}}}\right)^{K-1} \\
    &= {t_{\mathrm{set}}}-\dfrac{1}{K}\sum_{k=1}^{K-1}\binom{K}{k}B_k \dfrac{(t_{\mathrm{set}}-1)^{K-k}}{(t_{\mathrm{set}})^{K-1}}
\end{align}
where $B_k$ for $k={1,2,\ldots ,K-1}$ are the Bernoulli numbers\cite{Gould1972}. However, it is not easy to compute this value when $K$ is large. instead, we can derive a lower bound of $\mathbb{E}[t_K]$ based on $\mathbb{E}[\tilde{t}_K]$ by using the fact that $\tilde{t}_K \leq T_0t_K $ as follows:
\begin{align}
    \mathbb{E}[t_K]\geq\mathbb{E}\left[\dfrac{\tilde{t}_K}{T_0}\right] = \dfrac{\mathbb{E}[\tilde{t}_K]}{T_0}
     = \dfrac{K-1}{K}t_{\mathrm{set}}
\end{align}
Consequently, using this lower bound, we can derive the upper bound for $\mathbb{E}[{s_{\mathrm{min}}}]$ as 
\begin{align}
    \int_{d_K}&(1-\epsilon_K)T_0R_\text{UC} \left[t_{\mathrm{set}} - \mathbb{E}[t_K] \right]\mathrm{d}d_{K}\\
    &\leq \int_{d_K}(1-\epsilon_K)T_0R_\text{UC}\dfrac{t_{\mathrm{set}}}{K} \mathrm{d}d_{K}\\
    & \leq 
  \dfrac{t_{\mathrm{set}}}{K} T_0R_\text{UC}\exp\left(-\dfrac{2(2^{(\frac{R_\text{UC}}{w})}-1)D_\text{BS}^{\eta}}{(\eta+2)\rho_0  
  }\right)
\end{align}

\section{Proof of Proposition \ref{proposition4}} \label{appendix2}
\setcounter{equation}{0}
\renewcommand{\theequation}{B.\arabic{equation}}
The lower bound of the minimum cached data size among group users can be derived by order statistics. For convenience, we define the number of cached packets of user $k$ as $c_k=s_k/(T_0R_\text{UC})$ and the minimum number of the cached packets among $c_k$ for $\forall k$ as $c_{\mathrm{min}}$. Then, we can rewrite $\mathbb{E}[{s_{\mathrm{min}}}]$ as 
\begin{align}
    \mathbb{E}[{s_{\mathrm{min}}}] 
    &= T_0R_\text{UC} \cdot \mathbb{E}[\text{min}\{c_1,..., c_K \}]\\
    &=T_0 R_\text{UC} \sum_{c=0}^{t_{\mathrm{set}}}\text{Pr}[c_{\mathrm{min}}>c]\\
    &= T_0 R_\text{UC}\sum_{c=0}^{t_{\mathrm{set}}}(\Pi_{k=1}^{K}(1-\text{Pr}[c_k \leq c])) \label{appendixB1}
\end{align}
Now, we need to analyze the lower bound of \eqref{appendixB1}. For the first user, the remaining set-up phase time is determined and we can easily obtain the CDF of the number of cached packets in the set-up phase as
\begin{align}
    \text{Pr}[c_{1} \leq c]=\sum_{c'=0}^c {t_{\mathrm{set}} \choose c'} {\epsilon^{(t_{\mathrm{set}}-c')}_{1}}(1-\epsilon_{1})^{c'}
\end{align}  

For the rest group users, we can derive the CDF as follow.
\begin{align}
    \text{Pr}[c_{k} \leq c]
    =&\dfrac{c-1}{t_{\mathrm{set}}}\text{Pr}[c_{k} \leq c|t_{\mathrm{set}}-t_{k} < c] \nonumber\\
    &+ \dfrac{t_{\mathrm{set}}-c+1}{t_{\mathrm{set}}}\text{Pr}[c_{k} \leq c|t_{\mathrm{set}}-t_{k} \geq c].
\end{align}
If the remaining time slots are less than $c$, the minimum number of cached packets will be always less than $c$ and therefore,
\begin{align}
    \text{Pr}[c_{k} \leq c|t_{\mathrm{set}}-t_{k} < c]=1.
\end{align}
On the other hand, if the reamining time slots are equal to or exceed $c$, we can obtain the probability that the minimum number of cached packets is less than $c$ as

\begin{align}
    \text{Pr}&[c_{k} \leq c|t_{\mathrm{set}}-t_{k} \geq c] \nonumber\\
    &=\dfrac{\sum_{t_k=0}^{t_{\mathrm{set}}-c}\sum_{c'=0}^{c} {t_{\mathrm{set}}-t_{k} \choose c'} {\epsilon^{(t_{\mathrm{set}}-t_{k}-c')}_{k}}(1-\epsilon_{k})^{c'}}{t_{\mathrm{set}}-c+1}.
\end{align}
Substituting the above equations into  \eqref{appendixB1}, we can rearrange the equation as below.

\begin{align}
    \mathbb{E}&[{c_{\mathrm{min}}}]= \sum_{c=0}^{t_\text{set}}(\Pi_{k=1}^{K}(1-\text{Pr}[c_k \leq c]))\\
    & = \sum_{c=0}^{t_{\mathrm{set}}}\Bigg[\!\left(1\!-\!\sum_{c'=0}^c {t_{\mathrm{set}} \choose c'} {\epsilon^{(t_{\mathrm{set}}-c')}_{1}}(1-\epsilon_{1})^{c'}\right) \nonumber\\ 
    & \qquad\times \Pi_{k=2}^{K} \Bigg(\dfrac{c-1}{t_{\mathrm{set}}} +\sum_{t_k=0}^{t_{\mathrm{set}}-c}\sum_{c'=0}^{c} {t_{\mathrm{set}}-t_{k} \choose c'}\nonumber\\
    & \qquad\qquad\qquad\qquad \times \dfrac{{\epsilon^{(t_{\mathrm{set}}-t_{k}-c')}_{k}}(1-\epsilon_{k})^{c'}}{t_{\mathrm{set}}} \Bigg)\Bigg]\\
    & \geq \sum_{c=0}^{t_{\mathrm{set}}}\Bigg[\!\left(1\!-\!\sum_{c'=0}^c {t_{\mathrm{set}} \choose c'} {\epsilon^{(t_{\mathrm{set}}-c')}_{\max}}(1-\epsilon_{\max})^{c'}\right) \nonumber\\ 
    & \qquad\times \Pi_{k=2}^{K} \Bigg(\dfrac{c-1}{t_\text{set}} +\sum_{t_k=0}^{t_{\mathrm{set}}-c}\sum_{c'=0}^{c} {t_{\mathrm{set}}-t_{k} \choose c'}\nonumber\\
    & \quad\qquad\qquad\qquad \times \dfrac{{\epsilon^{(t_{\mathrm{set}}-t_{k}-c')}_{\max}}(1-\epsilon_{\max})^{c'}}{t_{\mathrm{set}}} \Bigg)\Bigg]
\end{align}
where $\epsilon_{\max}$ is the outage probability when the distance between the user and the BS is maximum (i.e., $D_\text{BS}$). To be specific, $\epsilon_{\max}$ can be written as
\begin{align}
    \epsilon_{\max}=1-\exp\left(-\dfrac{2^{(\frac{R_\text{C}}{w})}-1}{{\rho_0} D_\text{BS}^{-\eta}}\right).
\end{align}

\section{Proof of Proposition \ref{proposition5}} \label{appendix3}
\setcounter{equation}{0}
\renewcommand{\theequation}{C.\arabic{equation}}

Based on the definition of the minimum cached data size among the group users and under the assumption presented in Assumption \ref{assumption1},  we can rewrite the lower bound for $\mathbb{E}[{s_{\mathrm{min}}}]$ as 
\begin{align}
    \mathbb{E}&[{s_{\mathrm{min}}}] = \mathbb{E}[\text{min}\{s_1,..., s_K \}]\\
    &\geq\! \mathbb{E}\left[\mathrm{min} \{s_1 ,..., s_K \}|d_k \!=\! D_\text{BS}, ~ \forall  k\!\in\!\{1,...,K\}\right]\\
    &= \mathbb{E}[ s_K |d_K = D_\text{BS}]\\
    &=\!\sum_{t=1}^{t_{\mathrm{set}}}\!\Pr\left[t_K=t\right](t_{\mathrm{set}}-t)(1-\hat{\epsilon}_K)T_0 R_\text{UC} \\
    &=(1-\hat{\epsilon}_K)T_0 R_\text{UC} \left[t_{\mathrm{set}} - \mathbb{E}[t_K] \right]
\end{align}
Similar to Appendix \ref{appendix1}, we can derive the upper bound of $\mathbb{E}[t_K]$ based on $\mathbb{E}[\tilde{t}_K]$ by using the fact that $T_0t_K \leq \tilde{t}_K +T_0$.
\begin{align}
    \mathbb{E}[t_K]\leq\mathbb{E}\left[\dfrac{\tilde{t}_K}{T_0}+1\right] = \dfrac{\mathbb{E}[\tilde{t}_K]}{T_0} + 1 
     = \dfrac{K-1}{K}t_{\mathrm{set}} +1
\end{align}
Consequently, we can derive the lower bound for $\mathbb{E}[{s_{\mathrm{min}}}]$ as follows
\begin{align}
    \mathbb{E}&[{s_{\mathrm{min}}}]\geq(1-\hat{\epsilon}_K)T_0 R_\text{UC} \left[t_{\mathrm{set}} - \mathbb{E}[t_K] \right]\\
    &\geq
  \dfrac{t_{\mathrm{set}}-K}{K} T_0 R_\text{UC}\exp\left(-\dfrac{2^{(\frac{R_\text{UC}}{w})}-1}{{\rho_0} D_\text{BS}^{-\eta}}\right).
\end{align}

\IEEEpeerreviewmaketitle

\bibliographystyle{IEEEtran}  
\bibliography{IEEEabrv,Reference}

\end{document}